\documentclass[prl,showpacs,twocolumn,amsfonts,superscriptaddress]{revtex4}
\usepackage{graphicx}
\usepackage{color}
\usepackage{portland}

\begin{document}

\title{Temperature dependent electron-phonon coupling in a high T$_{c}$ cuprate probed by femtosecond X-ray diffraction.}

\author{B. Mansart}
\affiliation{Laboratory for Ultrafast Microscopy and Electron Scattering, ICMP, Ecole Polytechnique F\'{e}d\'{e}rale de Lausanne, CH-1015 Lausanne, Switzerland}
\author{M.J.G. Cottet}
\affiliation{Laboratory for Ultrafast Microscopy and Electron Scattering, ICMP, Ecole Polytechnique F\'{e}d\'{e}rale de Lausanne, CH-1015 Lausanne, Switzerland}
\author{G.F. Mancini}
\affiliation{Laboratory for Ultrafast Microscopy and Electron Scattering, ICMP, Ecole Polytechnique F\'{e}d\'{e}rale de Lausanne, CH-1015 Lausanne, Switzerland}
\author{T. Jarlborg}
\affiliation{DPMC, University of Geneva, 24 Quai Ernest-Ansermet, CH-1211 Geneva 4, Switzerland}
\author{S.B. Dugdale}
\affiliation{H. H. Wills Physics Laboratory, University of Bristol, Tyndall Avenue, Bristol BS8 1TL, UK}
\author{S.L. Johnson}
\affiliation{Swiss Light Source, Paul Scherrer Institut, CH-5232 Villigen, Switzerland}
\affiliation{ETH-Z\"urich, CH-8093 Z\"urich, Switzerland}
\author{S.O. Mariager}
\affiliation{Swiss Light Source, Paul Scherrer Institut, CH-5232 Villigen, Switzerland}
\author{C.J. Milne}
\affiliation{Laboratory of Ultrafast Spectroscopy, ISIC, Ecole Polytechnique F\'{e}d\'{e}rale de Lausanne, CH-1015 Lausanne, Switzerland}
\author{P. Beaud}
\affiliation{Swiss Light Source, Paul Scherrer Institut, CH-5232 Villigen, Switzerland}
\author{S. Gr\"ubel}
\affiliation{Swiss Light Source, Paul Scherrer Institut, CH-5232 Villigen, Switzerland}
\author{J.A. Johnson}
\affiliation{Swiss Light Source, Paul Scherrer Institut, CH-5232 Villigen, Switzerland}
\author{T. Kubacka}
\affiliation{ETH-Z\"urich, CH-8093 Z\"urich, Switzerland}
\author{G. Ingold}
\affiliation{Swiss Light Source, Paul Scherrer Institut, CH-5232 Villigen, Switzerland}
\author{K. Prsa}
\affiliation{Laboratory for Quantum Magnetism, ICMP, Ecole Polytechnique F\'{e}d\'{e}rale de Lausanne, CH-1015 Lausanne, Switzerland}
\author{H.M. R\o{}nnow}
\affiliation{Laboratory for Quantum Magnetism, ICMP, Ecole Polytechnique F\'{e}d\'{e}rale de Lausanne, CH-1015 Lausanne, Switzerland}
\author{K. Conder}
\affiliation{Laboratory for Developments and Methods, PSI, CH-5232 Villigen PSI, Switzerland}
\author{E. Pomjakushina}
\affiliation{Laboratory for Developments and Methods, PSI, CH-5232 Villigen PSI, Switzerland}
\author{M. Chergui}
\affiliation{Laboratory of Ultrafast Spectroscopy, ISIC, Ecole Polytechnique F\'{e}d\'{e}rale de Lausanne, CH-1015 Lausanne, Switzerland}
\author{F. Carbone}
\affiliation{Laboratory for Ultrafast Microscopy and Electron Scattering, ICMP, Ecole Polytechnique F\'{e}d\'{e}rale de Lausanne, CH-1015 Lausanne, Switzerland}

\begin{abstract}

The strength of the electron-phonon coupling parameter and its evolution throughout a solid's phase diagram often determines phenomena such as superconductivity, charge- and spin-density waves. Its experimental determination relies on the ability to distinguish thermally activated phonons from those emitted by conduction band electrons, which can be achieved in an elegant way by ultrafast techniques. Separating the electronic from the out-of-equilibrium lattice subsystems, we probed their re-equilibration by monitoring the transient lattice temperature through femtosecond X-ray diffraction in La$_{2-x}$Sr$_x$CuO$_4$ single crystals with $x$=0.1 and 0.21. The temperature dependence of the electron-phonon coupling is obtained experimentally and shows similar trends to what is expected from the \textit{ab-initio} calculated shape of the electronic density-of-states near the Fermi energy. This study evidences the important role of band effects in the electron-lattice interaction in solids, in particular in superconductors. 

\end{abstract}


\maketitle

\section{INTRODUCTION}

Electron-phonon (e-ph) coupling is a key parameter for describing the properties of solids. It is particularly important for superconductors, since it mediates the electron pairing in its conventional form, described by the Bardeen-Cooper-Schrieffer theory~\cite{BCS}. On the other hand, even though many attempts have been made to account for the high critical temperatures observed in cuprates, e-ph coupling seems unable to provide the unconventional superconductivity mechanism even in the strong coupling regime~\cite{Scalapino1966}. Nevertheless, the peculiar density-of-states (DOS) and Fermi surface of the cuprates reveal interesting properties related to e-ph coupling~\cite{Devereaux2004, Carbone2008}, which undoubtedly play a role in the evolution of their electronic properties throughout the phase diagram.

In pump-probe experiments, intense fs light pulses induce a rapid jump in the electronic temperature of the material (the out-of-equilibrium electron distribution typically thermalizing within a few tens of fs), followed by a slower ($\sim$ps) re-equilibration with the lattice temperature through energy transfer via e-ph coupling~\cite{AllenStatic, allen}. 
 
The transient electronic temperature can be directly measured by PhotoElectron Spectroscopy~\cite{Fann1992} and optics~\cite{Mansart2012}, while the transient lattice temperature can be obtained via diffraction~\cite{Gedik2007, Carbone2008}. The relaxation of these observables can be described by a multi-temperature model which in turns yields the e-ph coupling parameter~\cite{Carbone2008, allen, Perfetti2007, Mansart2012, Mansart2010, Kaganov1957}, and in the case of $k$-sensitive probes like diffraction~\cite{Carbone2008} and Angle-Resolved PhotoElectron Spectroscopy (ARPES)~\cite{Perfetti2007, Cortes2011}, its symmetry as well.

In this paper, we present a combined theoretical and experimental study of the e-ph coupling in La$_{2-x}$Sr$_x$CuO$_4$ (LSCO). Calculating the energy distribution of the DOS for different electronic temperatures $T$, we demonstrate that the e-ph coupling can depend on $T$, even when the latter reaches very high values. This effect is verified by means of time-resolved X-ray diffraction for different Sr dopings, showing the evolution of the e-ph interactions in the phase diagram of a cuprate superconductor.

Time-resolved X-ray diffraction experiments were performed using the FEMTO slicing source located at the MicroXAS beamline of the Swiss Light Source (Paul Scherrer Institute). After excitation with 1.55 eV photons, we probed the transient lattice temperature by measuring (in an asymmetric scattering geometry~\cite{Johnson2010}) the (400) Bragg peak, corresponding to the antinodal direction, of two La$_{2-x}$Sr$_x$CuO$_4$ single crystals ($x=0.1$ and $x=0.21$). 
In cuprate systems, the latter corresponds to the strongest e-ph coupling coming from the interaction between antinodal carriers and specific in-plane lattice modes~\cite{Devereaux2004, Carbone2008}.
The X-ray source delivers 200 photons per pulse at a 2 kHz repetition rate; its energy was varied between 7.5 and 8 keV, and  its incidence angle was chosen to be 0.87$^{\circ}$ for both samples in order for the pump and the probe penetration depths to coincide. The overall time-resolution was 200 fs~\cite{Beaud2007}. The pump beam had a duration of 100 fs and fluences ranging from 5 to 27.2 $mJ/cm^2$; all measurements were performed at room temperature. 
A schematic of the experimental setup can be found in Refs.\cite{Beaud2007} and~\cite{Beaud2011}.

\section{TIME-RESOLVED X-RAY DIFFRACTION DATA}

\begin{figure}[h]
\includegraphics[width=1\linewidth,clip=true]{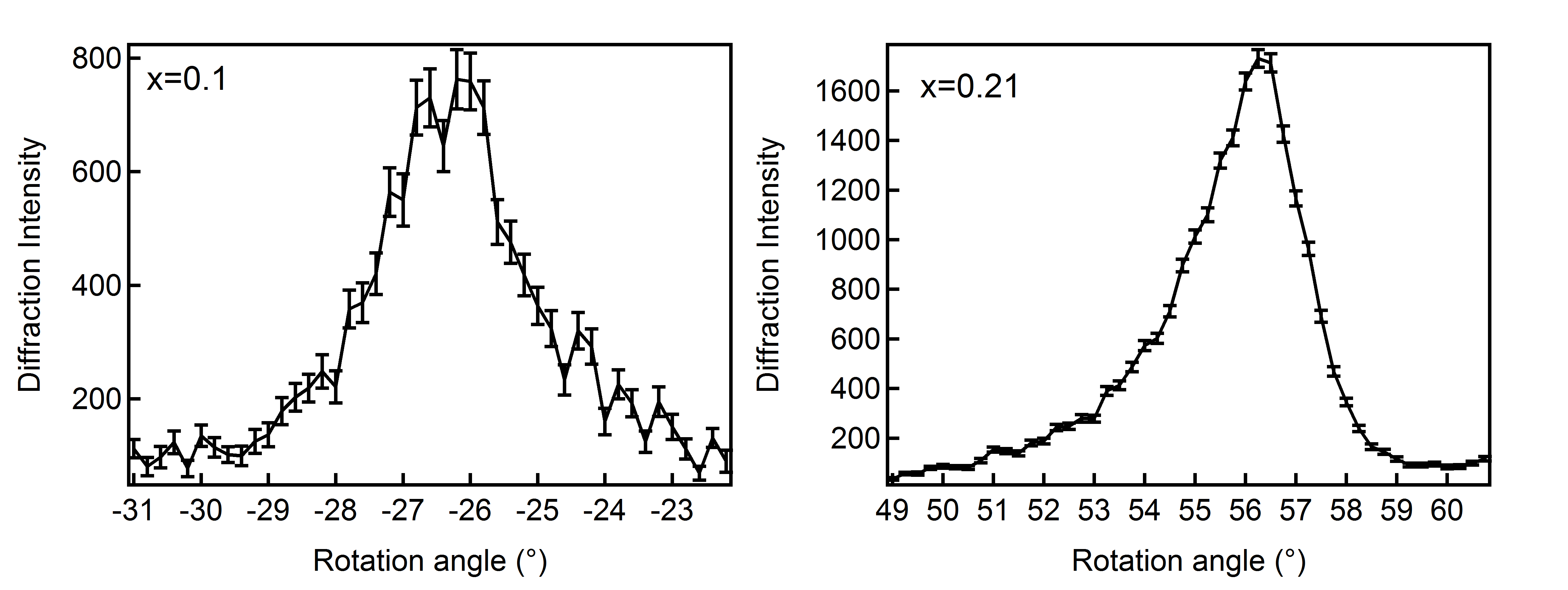}
\caption{Rocking curves obtained for the La$_{2-x}$Sr$_x$CuO$_4$ samples with the core beam of Micro-XAS-FEMTO beamline.}
\label{rock_static}
\end{figure}

The first step before measuring a pump-probe signal on a diffraction peak is to find its position in asymmetric geometry, since the X-ray incidence angle has to be kept grazing. The rocking curves corresponding to the (400) peak are shown in Fig.~\ref{rock_static}. The sample orientations were (230) for $x=0.1$ and (211) for $x=0.21$.

We checked carefully the behavior of these rocking curves as a function of the time delay. Indeed, a transient temperature analysis can be performed only if the structural properties remain the same as the unperturbed compound, meaning that the lattice is not thermally distorted. After a thermal dilatation, the system is too different from the initial state to obtain meaningful information about the compound at equilibrium.

\begin{figure}[h!]
\includegraphics[width=1\linewidth,clip=true]{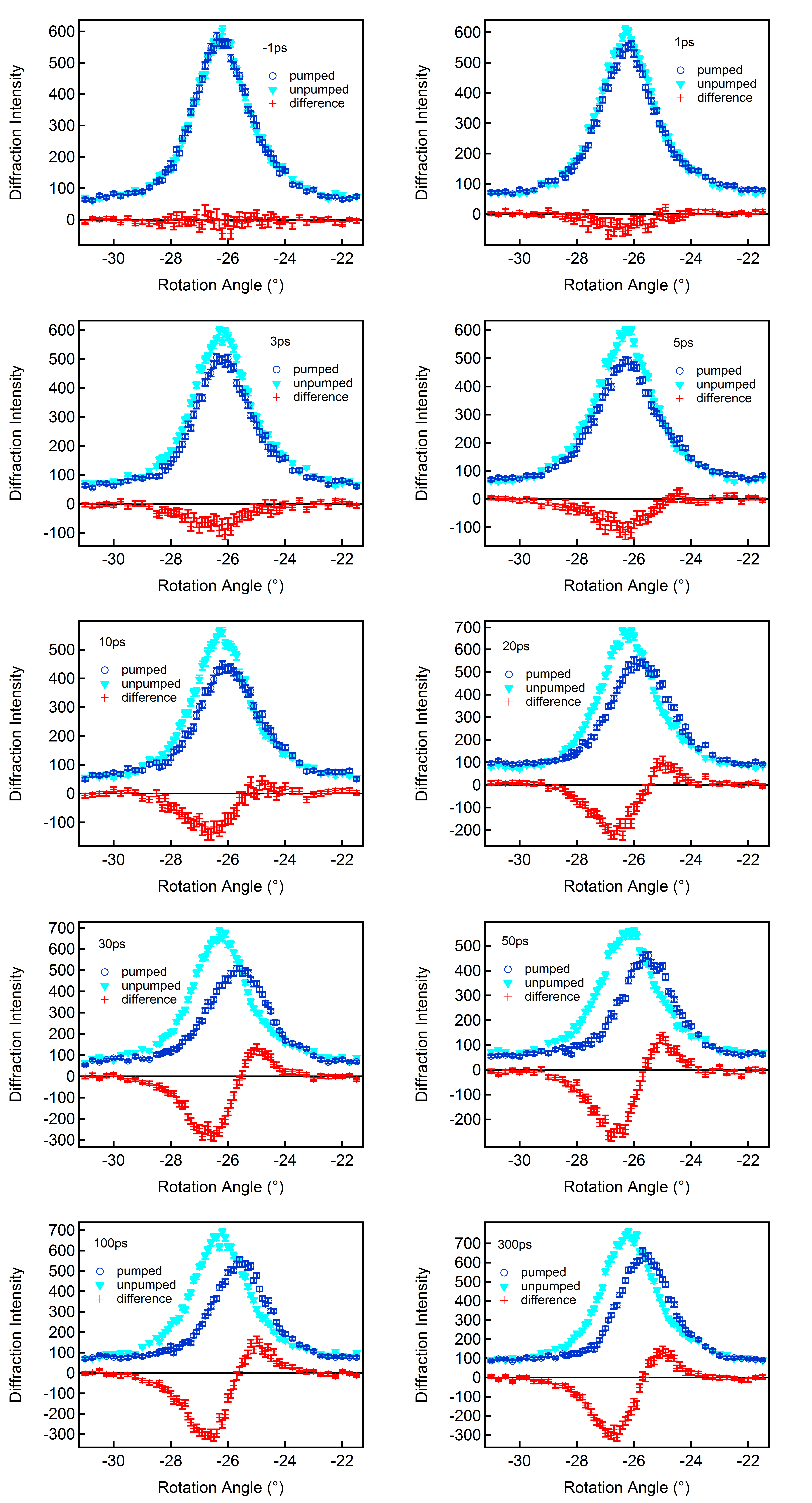}
\caption{Rocking curve of the (400) peak in La$_{2-x}$Sr$_x$CuO$_4$, $x=0.1$, for different time delays after excitation. Light blue triangles are the un-pumped curve, dark blue circles the pumped one and the difference between them is shown as red crosses.}
\label{rotscans_x01}
\end{figure}

\begin{figure}[h!]
\includegraphics[width=1\linewidth,clip=true]{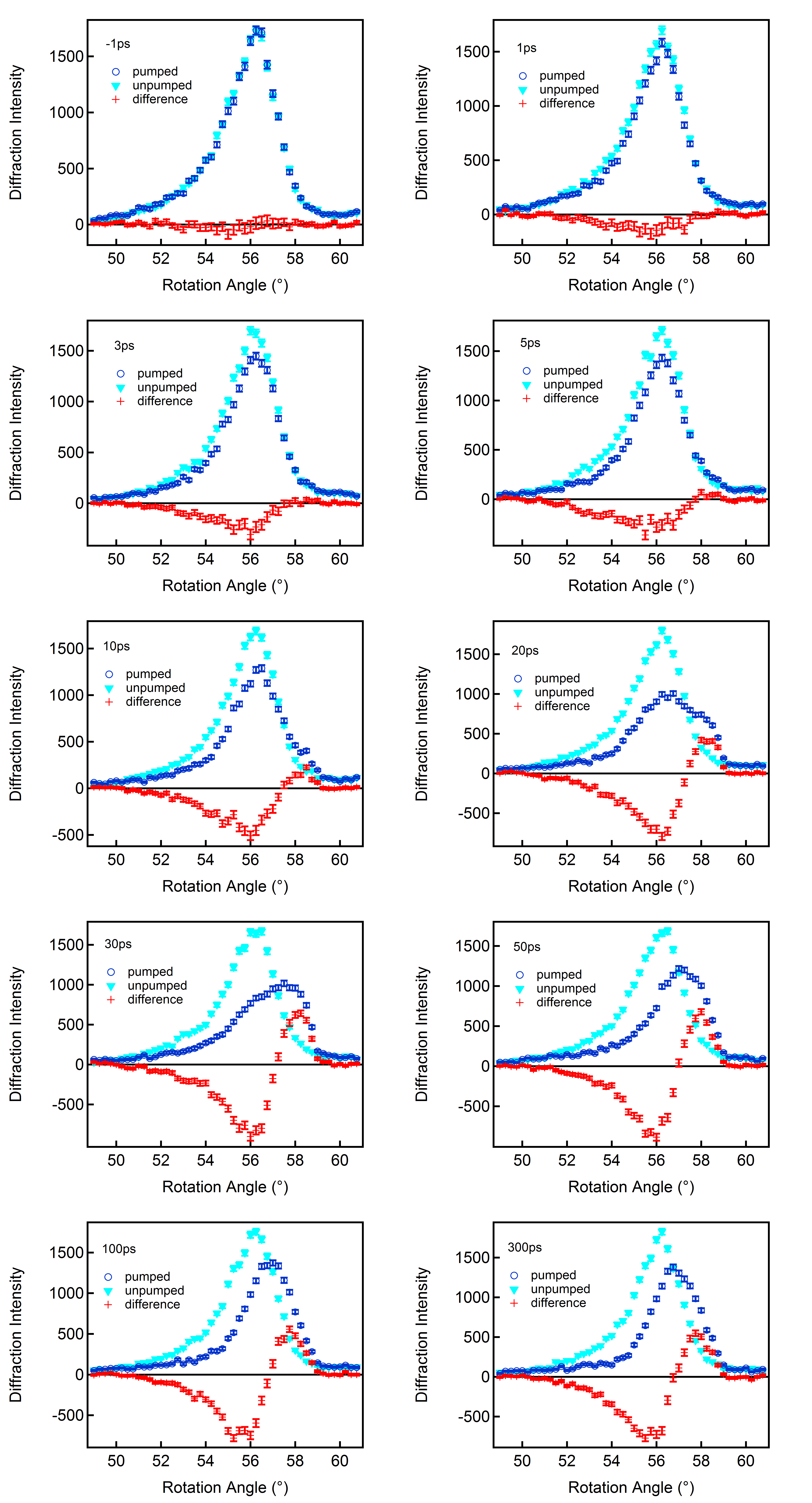}
\caption{Rocking curve of the (400) peak in La$_{2-x}$Sr$_x$CuO$_4$, $x=0.21$, for different time delays after excitation. Light blue triangles are the un-pumped curve, dark blue the pumped one and the difference between them is shown as red crosses.}
\label{rotscans_x021}
\end{figure}

This dilatation is evidenced by the Bragg peak shifting towards larger diffraction angles. The rocking curves of excited and non-excited systems are presented in Figs.~\ref{rotscans_x01} ($x=0.1$) and~\ref{rotscans_x021} ($x=0.21$) for a pump fluence of 20.5 $mJ/cm^2$, from which we can see a peak shift occuring between 5 and 10 ps after excitation. The system can therefore be considered as being slightly perturbed only during the first 5 ps.

This thermal dilatation is due to the heat transport by acoustic waves after photoexcitation towards the bulk of the material. Indeed, the time needed for the longitudinal acoustic phonons of speed $v_s\approx 4000~cm/s$~\cite{Boni1988} to propagate across the penetration depth distance of $l=60~nm$~\cite{calc} is $t=l/v_s\approx 15~ps$, close to the value experimentally found.

There is a striking difference in the peak lineshape for the two studied dopings, particularly visible for delays in the 20-50 ps range. In the overdoped sample, the Bragg peak seems to be formed by two different peaks, as expected in a compound containing phase separation between two domains having slightly different lattice parameters. This behavior is possibly reminiscent of what is observed in La$_2$CuO$_{4+ \delta}$ thin films by ultrafast electron diffraction~\cite{Gedik2007}, but it could also be due to a slight misalignment of the detector on the diffraction peak. This possibility does not spoil the data analysis and interpretation.

\begin{figure}[h!]
\includegraphics[width=1\linewidth,clip=true]{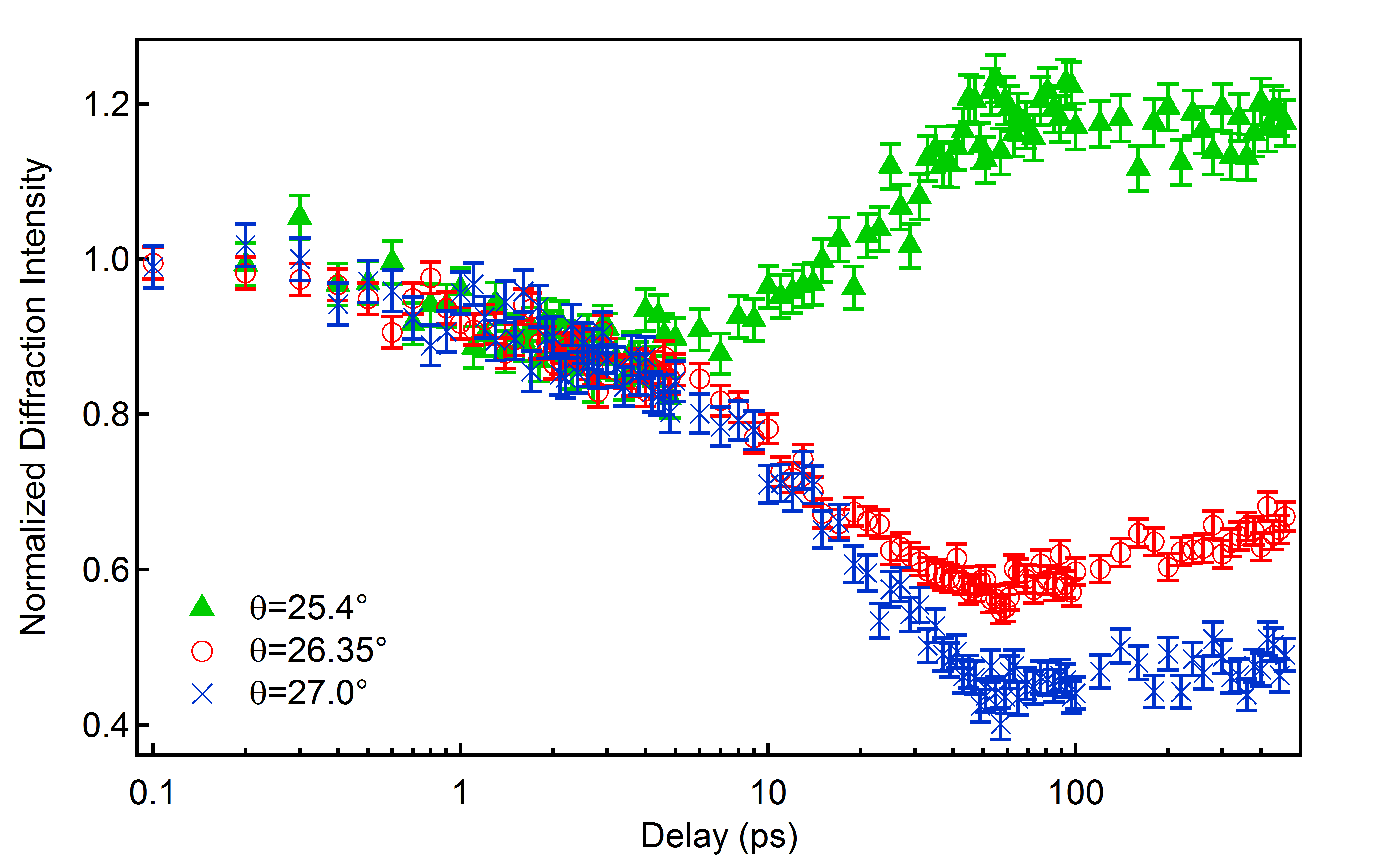}
\caption{Delay scans at different angles of the rocking curve of the (400) peak in La$_{2-x}$Sr$_x$CuO$_4$, $x=0.1$.}
\label{Longdelayscans_x01}
\end{figure}

\begin{figure}[h!]
\includegraphics[width=1\linewidth,clip=true]{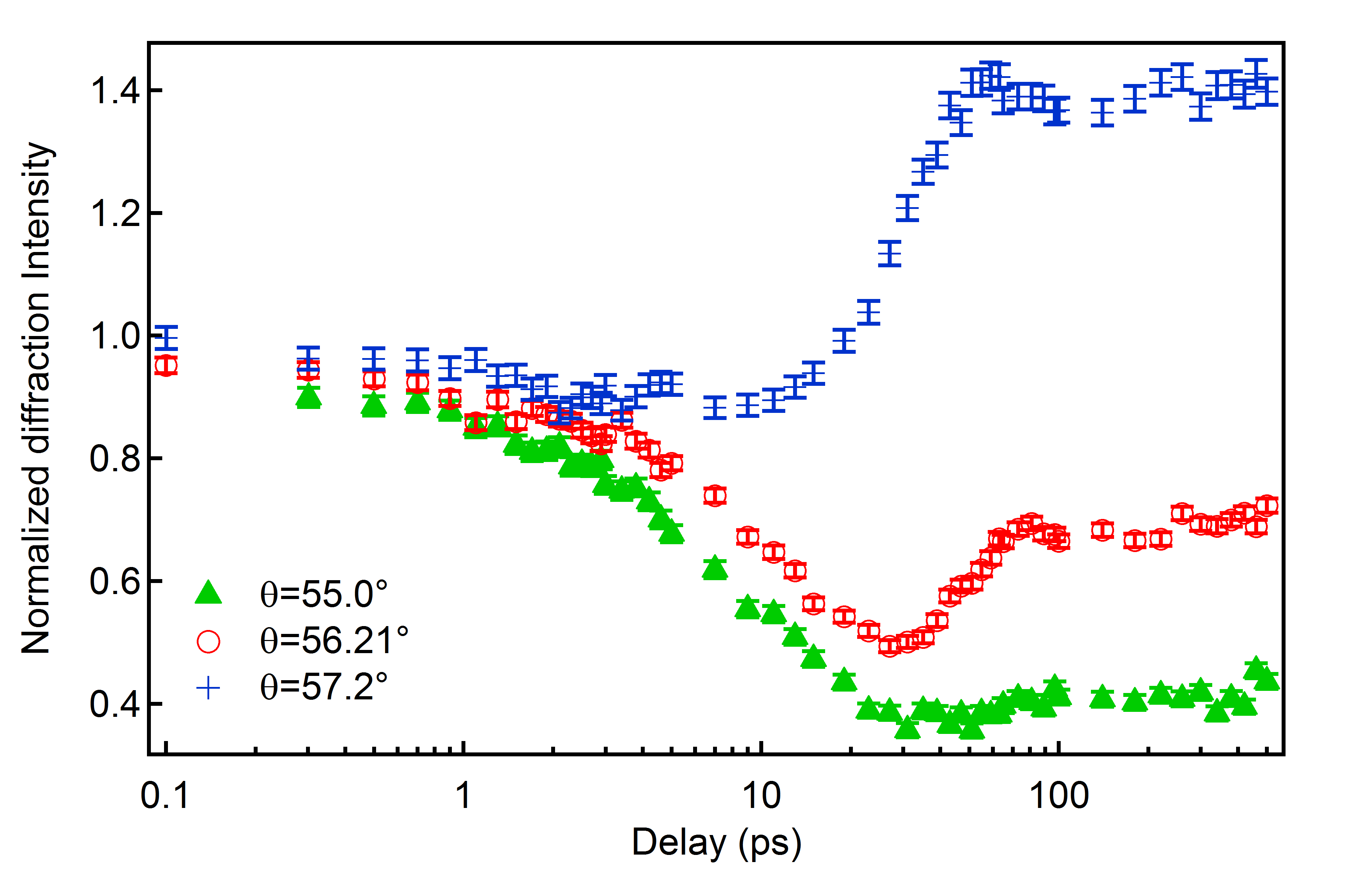}
\caption{Delay scans at different angles of the rocking curve of the (400) peak in La$_{2-x}$Sr$_x$CuO$_4$, $x=0.21$.}
\label{Longdelayscans_x021}
\end{figure}

In order to follow the time-dependence of the Bragg peak intensity over a time range of several hundred picoseconds, long delay scans have been performed for three different diffraction angles. The latter correspond to the center of the unperturbed peak, and one larger and one smaller diffraction angle. These long delay scans are shown in Figs.~\ref{Longdelayscans_x01} and \ref{Longdelayscans_x021}. From these results, one can conclude that in the first 5 ps after excitation, no lattice dilatation occurs; indeed, the diffraction intensity does not depend on the angle, which indicates that the peak shift towards larger angles is negligible.

The peak shift starts around 5 ps, where the time-dependent diffraction intensity behavior changes between lower and larger angles. We deduced from this observation that performing a transient temperature analysis was correct if one considers only the first 5 ps of the measurement.

\section{TRANSIENT LATTICE TEMPERATURE}

\begin{figure}[ht]
\includegraphics[width=0.8\linewidth,clip=true]{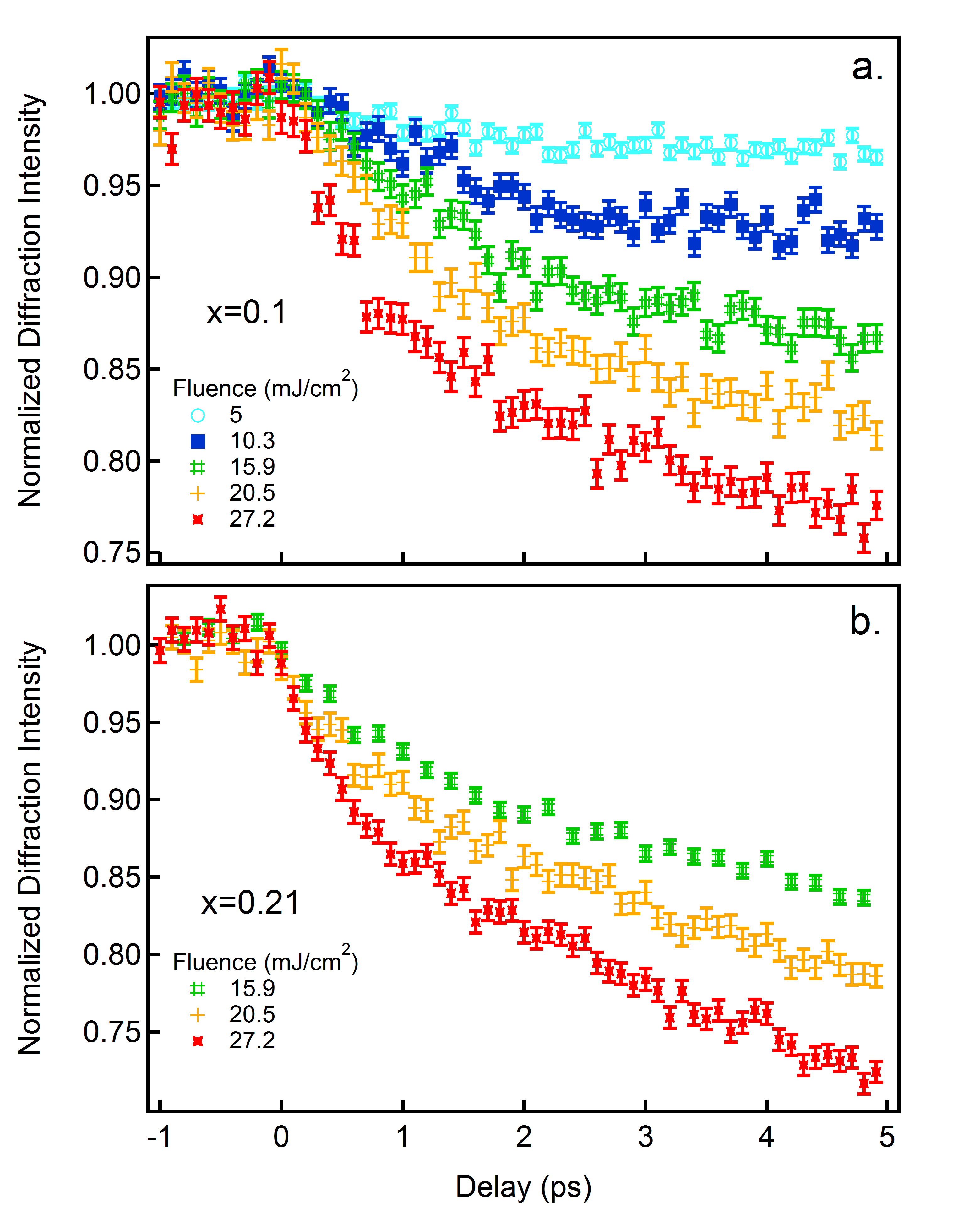}
\caption{Diffraction intensity as a function of time delay for the (400) peak in La$_{2-x}$Sr$_x$CuO$_4$, $x=0.1$ (a) and $x=0.21$ (b). The diffraction angles are set at -26.35$^{\circ}$ ($x=0.1$) and 56.21$^{\circ}$ ($x=0.21$) i.e. in the center of the Bragg peak.}
\label{fig1}
\end{figure}

The temporal evolution of the normalized diffraction intensities of the (400) Bragg diffraction peak, measured as the ratio between the pumped and unpumped signals, are presented in Fig.~\ref{fig1} for the two different dopings studied. 
The Bragg peak intensity was checked not to change significantly before
and after the pump pulse impinged on the sample, indicating that no significant average heating took place at this repetition-rate. Also, the flatness of the baseline before time zero indicates that the system fully recovers its equilibrium condition between pulses (see Fig.~\ref{fig1}).

\begin{figure}[ht]
\includegraphics[width=0.8\linewidth,clip=true]{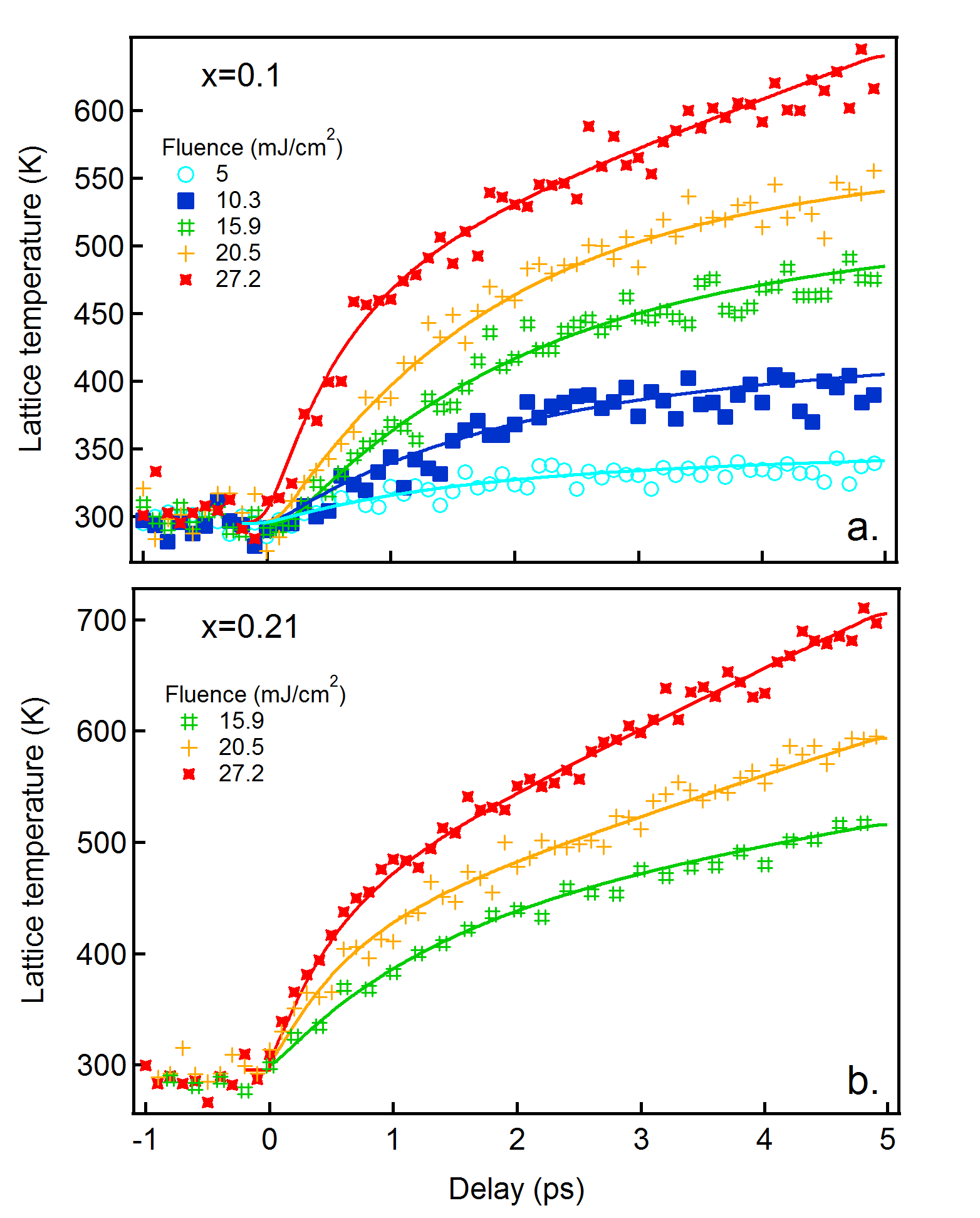}
\caption{Transient lattice temperature obtained from the (400) diffraction peak intensity of LSCO, $x=0.1$ (a) and $x=0.21$ (b). Markers represent experimental data and solid lines the corresponding 3TM simulations, from which we plot the effective average lattice temperature as $T_L=\alpha~T_h+ (1-\alpha)~T_c$.}
\label{fig2}
\end{figure}

We assume that the lattice can be described as an ensemble of different sets of phonons, each of them separately in thermal equilibrium, in which the effective average temperature of the lattice increases due to the energy exchange with heated electrons. The diffraction intensity is then directly related to the effective average lattice temperature through the Debye model for a non-distorted lattice. A decrease of the initial (equilibrium) Bragg diffraction intensity indicates the population of phonons which spoil the diffraction condition by disordering the interatomic distances. The average atomic displacements induced by such phonons can be evaluated by comparing the perturbed and unperturbed diffraction intensities denoted by $I(t)$ and $I_0$, respectively. The latter is the diffraction intensity at the initial temperature, in our case at $T_0=295~K$.

We consider the following expression for diffracted intensity in the presence of atomic disorder:

\begin{equation}
I(\textbf{q})=N^2 \left|\left\langle F_n (\textbf{q})\right|\right\rangle^2= N^2 f^2 exp(-2W)
\end{equation}

Where $F_n$ is the structure factor, $N$ the number of unit cells; \textbf{q} is the reciprocal lattice vector corresponding to the measured Bragg peak, so for (400) in La$_{2-x}$Sr$_x$CuO$_4$, $q=6.65~\AA^{-1}$. The term $exp(-2W)$ is the Debye-Waller factor defined in the presence of an atomic displacement $\textbf{u}_n$ as:

\begin{equation}
W=\frac{1}{2} \left\langle (\textbf{q}.\textbf{u}_n)^2\right\rangle
\end{equation}

The atomic motions considered in this formula are due to the finite lattice temperature. Therefore, considering an isotropic average for these displacements, one can relate $\left\langle u^2\right\rangle$ to the lattice temperature by the Debye formula:

\begin{equation}
\left\langle u^2\right\rangle=\frac{9\hbar^2 T_L}{M k_B \Theta_D^2}
\end{equation}

with $T_L$ the lattice temperature, $M$ the mass of one unit cell and $\Theta_D$ the Debye temperature. 

In the case of a time-dependent experiment, one needs to consider an increase in the lattice temperature, inducing (as a function of time) an increase in thermal agitation which reduces the diffraction intensity. Then one compares the perturbed and the unperturbed values for diffracted intensity. The expression for calculating the transient lattice temperature is:

\begin{equation}
T_L (t)=T_0-\frac{M k_B \Theta_D^2}{3 \hbar^2 q^2}\ln\left(\frac{I(t)}{I_0}\right)
\end{equation}

 To obtain the Debye temperature accurately, we measured the high-temperature specific heat of LSCO as a function of temperature, from which we obtained a value of $\Theta_D~=~377~K$ (see the following section). 
The transient effective average lattice temperatures extracted from this analysis are shown in Fig.~\ref{fig2}. 

\section{HIGH-TEMPERATURE SPECIFIC HEAT MEASUREMENTS}
\label{Cp}

\begin{figure}[h!]
\includegraphics[width=1\linewidth,clip=true]{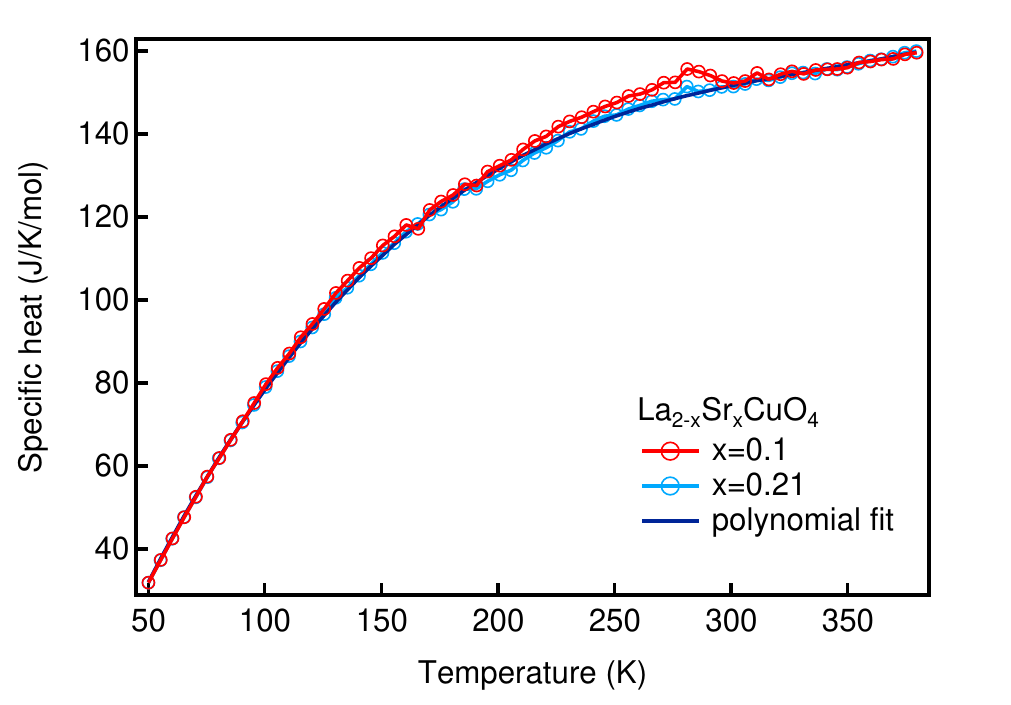}
\caption{Specific heat measurements in La$_{2-x}$Sr$_x$CuO$_4$, $x=0.1$ and $x=0.21$.}
\label{specific_heat}
\end{figure}

In order to calculate the lattice temperatures and to perform Three-Temperature Model simulations, we needed the lattice part of the specific heat up to the Debye temperature $\Theta_D$. We performed therefore measurements of the heat capacity for our samples at the Paul Scherrer Institute (Villigen, Switzerland).

We measured small samples of the same batch as those used for time-dependent X-ray diffraction, having masses of $17.3~mg$ for $x=0.1$ and $19.1~mg$ for $x=0.21$.

The $C_p(T)$ measurements started with measuring the contribution of the small amount of H-grease (used to provide thermal contact with the samples) in the range $50~K$-$380~K$.

After rescaling the results in order to take into account the H-grease contribution, we obtained the specific heat shown in Fig.~\ref{specific_heat}, together with a polynomial fit used to extract the lattice specific heat at all temperatures above $T_0$. The Debye temperature obtained from these data is $\Theta_D=377~K$, and we verified that at these high temperatures with respect to $T_c$, both dopings presented an identical behavior (the noise in the $x=0.1$ sample data is an experimental artifact). 
This value is in reasonable agreement with previously reported values, $\Theta_D\approx 420-450~K$~\cite{Momono}, leading to a maximum error of 19$\%$. Such an uncertainty has a very weak influence on our results, as well as on our qualitative conclusions.

In the following, we use the specific heat values in volume units rather than molar ones, in order to perform simulations with a depth-dependent model such as the Three-Temperature Model described in the next section.

\section{THREE-TEMPERATURE MODEL SIMULATIONS}

From the transient effective average lattice temperatures one can access the e-ph coupling constant through Three-Temperature Model (3TM) simulations. Indeed, while the Two-Temperature Model was first 
introduced~\cite{Kaganov1957, allen} in order to describe the energy transfer between the electron and lattice subsystems, its validity relies on the electronic temperature being larger than the Debye temperature and on the isotropy of the e-ph coupling function. For anisotropic materials such as cuprates~\cite{Perfetti2007}, iron-pnictides~\cite{Mansart2010} or charge-density waves systems~\cite{Mansart2012}, or in the case when only one diffraction peak is measured, a selective coupling between electrons and a subset of the total phonon modes may be taken into account using the 3TM, governed by the following equations:

\begin{eqnarray}
\label{eq_3TM}
2 C_e \frac{\partial T_e}{\partial t} &=& \frac{2(1-R)}{l_s}I_L(t)-g(T_e-T_h) \nonumber \\
\alpha C_L \frac{\partial T_h}{\partial t} &=& g(T_e-T_h)-g_c(T_h-T_c)\\
(1-\alpha)C_{L} \frac{\partial T_c}{\partial t} &=& g_c(T_h-T_c) \nonumber\\
\end{eqnarray}

where $T_e$, $T_h$ and $T_c$ are the temperatures of the electrons, the efficiently coupled (hot) phonons and the remaining modes, respectively. $C_e$=$\gamma T_e$ is the electronic specific heat ($\gamma$=158~$J\cdot m^{-3}\cdot K^{-2}$~\cite{Wada1989}), and $C_L$ is that of the lattice (taken from our own measurements). The calculated $\gamma$ from the bare (low temperature) DOS is about 70 $J \cdot m^{-3}\cdot K^{-2}$, which leaves room for a large e-ph coupling constant from phonons and/or spin fluctuations~\cite{tj09}. $\alpha$ is the fraction of efficiently coupled modes, $R$ the static reflectivity ($R=0.22$ for 1.55 eV in $p$-polarization arriving at 10$^{\circ}$ from the surface) and $l_s$ the penetration depth ($l_s=206~nm$ at 1.55 eV), both taken at the pump energy, and $I_L(t)$ is the pump intensity. The constant $g$ governs the energy transfer rate from electrons to hot phonons, and is related to the second moment of the Eliashberg function $\lambda \left\langle \omega^2 \right\rangle$ through $g=\frac{6\hbar\gamma}{\pi k_B} \lambda \left\langle \omega^2 \right\rangle$~\cite{allen}, $\lambda$ being the dimensionless e-ph coupling constant, whose strength averages over the interactions between many different electronic and phonon states. $g_c$ is the anharmonic coupling parameter which controls the energy relaxation from coupled phonons to the rest of the lattice.
Noteworthy, those parameters are rather independent from each other, which enhance our confidence on the results of these simulations.

We performed 3TM simulations assuming temperature-independent parameters; indeed, even though $\gamma$, $\Theta_D$ and $\lambda$ may depend on temperature~\cite{Lin2008, Johnson2009}, calculating their anharmonicities would be speculative in the case of room-temperature LSCO. Therefore we used the experimental values determined at equilibrium for $\gamma$ and $\Theta_D$ and assumed a constant $\lambda$ parameter over time (and therefore over temperature) for each excitation fluence. We used an iterative procedure, calculating at each time step the depth-dependent temperature profiles. At each depth and time step, we iterate the electronic and lattice part of the specific heat. This procedure is detailed in the Supporting Information of Ref.~\cite{Mansart2012}.
The depth used to calculate the temperatures is the X-ray penetration depth (since we calculate the measured temperatures), $l=60~nm$~\cite{calc}.

Coupling between electrons and electronic excitations (such as spin-fluctuations) are excluded from our 3TM simulations, as well as coupling between phonons and such excitations which might be important for underdoped LSCO~\cite{tj09, tjapl09, tjprb11}. We point out that these kind of relaxation processes may exist, but are not reachable by diffraction techniques which allow access only to the lattice temperature, and not the electronic one. The importance of spin-fluctuations in the bosonic glue function has been determined by static optical spectroscopy~\cite{vanHeumen2009}, which found them to be relevant in the high-energy excitation region (up to 300 meV), whereas phonons are limited to a lower energy-range (around 50-60 meV). As far as time-scale is concerned, it would suggest that coupling between electrons and spin-fluctuations is faster than that between electrons and phonons, as found by time-resolved spectroscopy~\cite{Dalconte2012}.

Neglecting the possibility that spin fluctuations could be preferentially excited by hot electrons rather than phonons implies that we overestimated the number and temperature of electrons in our model, since the latter takes only into account electrons thermalizing with phonons. Therefore, this omission would result in the absolute strength of electron-phonon coupling being somewhat underestimated, without affecting the trends nor our conclusions.

On the other hand, there may be couplings between phonons and spin-fluctuations, especially in the underdoped part of the phase diagram (see for example Ref.~\cite{tj09, tjapl09, tjprb11}). This would imply that our 3TM simulations overestimated the anharmonic coupling parameter of our model ($g_c$), and it would not affect our main conclusions either.

\begin{table}[!h]
\begin {center}
\begin{tabular}{|c|c|c|c|c|c|c|}
\hline
$x$ & $F$  & $T_e$ &   $\alpha$ & $g$ ($\times 10^{17}$) & $\lambda \left\langle \omega^2 \right\rangle$ & $\lambda$\\
~ & $(mJ/cm^2)$ & $(K)$  & & $(J/m^3/s/K)$  &  ($meV^2$) &  \\
\hline
  & 5 & 1430 & 0.05 & $0.7$ & 13.1 & 0.043  \\
  & 10.3 & 2030 & 0.085 & $1.0$ & 18.8 &  0.063  \\
 0.1 & 15.9 & 2513 & 0.14 & $1.3$ & 24.4 & 0.082   \\
  & 20.5 & 2849 & 0.15 & $1.6$ & 30.0 & 0.101  \\
    & 27.2 & 3278 &  0.08 & $3.0$ & 56.3 & 0.187  \\
\hline
  & 15.9 & 2513 &  0.1 & $1.6$ & 30.0 & 0.094   \\
0.21  & 20.5 & 2849 & 0.065 & $2.7$ & 50.7 & 0.158   \\
  & 27.2 & 3278 & 0.065 & $3.35$ & 62.9 & 0.196  \\
\hline
		\end{tabular}
	\caption{Maximum electronic temperature, fraction of coupled modes and electron-phonon coupling constants extracted from 3TM simulations.}
\label{3TM_results}
\end {center}
\end{table}

The simulations corresponding to the transient effective average lattice temperature are shown in Fig.~\ref{fig2}, the model is found to reproduce the experimentally derived lattice temperature very well for both doping levels and every pumping fluence.

The e-ph coupling constant $\lambda$ was obtained via these simulations, given average phonon energies $\left\langle \omega \right\rangle$ of 17.32 meV for $x=0.1$ and 17.89 meV for $x=0.21$. This average takes into account only the modes involving atomic motion along the $a$-axis~\cite{Mostoller1990}, which are mainly influencing the intensity of the (400) diffraction peak, and having a finite $\Gamma$-point e-ph coupling constants as calculated at $T=0~K$ in the QUANTUM ESPRESSO code~\cite{qe}. It is noteworthy that these calculations provide a value of $\lambda=0.031$ ($x=0.1$) and $0.029$ ($x=0.21$), in reasonable agreement with the 3TM results at the lowest fluences (see below).

The results of the 3TM simulations are given in Table~\ref{3TM_results}. We obtained the e-ph coupling constant $\lambda$ and the fraction of efficiently coupled modes $\alpha$. The values obtained for $\lambda$ are smaller than those found by means of $k$-integrated probes such as optics~\cite{Mansart2010, Mansart2012}, this may be because our experiments only probe a fraction of the whole phonon bath.

For each excitation fluence, the system reaches a given electronic temperature in the skin depth at initial time. This electronic temperature may be calculated through the formula:

\begin{equation}
T_e=\left\langle\sqrt{T_0^2+\frac{2(1-R)F}{l_s \gamma}e^{-z/l_s}}\right\rangle
\end{equation}

where the average is taken over the penetration depth of the pump pulses ($l_s$), and $F$ is the pumping fluence. The values of $T_e$ are also reported in Table~\ref{3TM_results}.

\section{TEMPERATURE DEPENDENCE OF THE DENSITY OF STATES}

 \begin{figure}[h!]
\includegraphics[width=1\linewidth,clip=true]{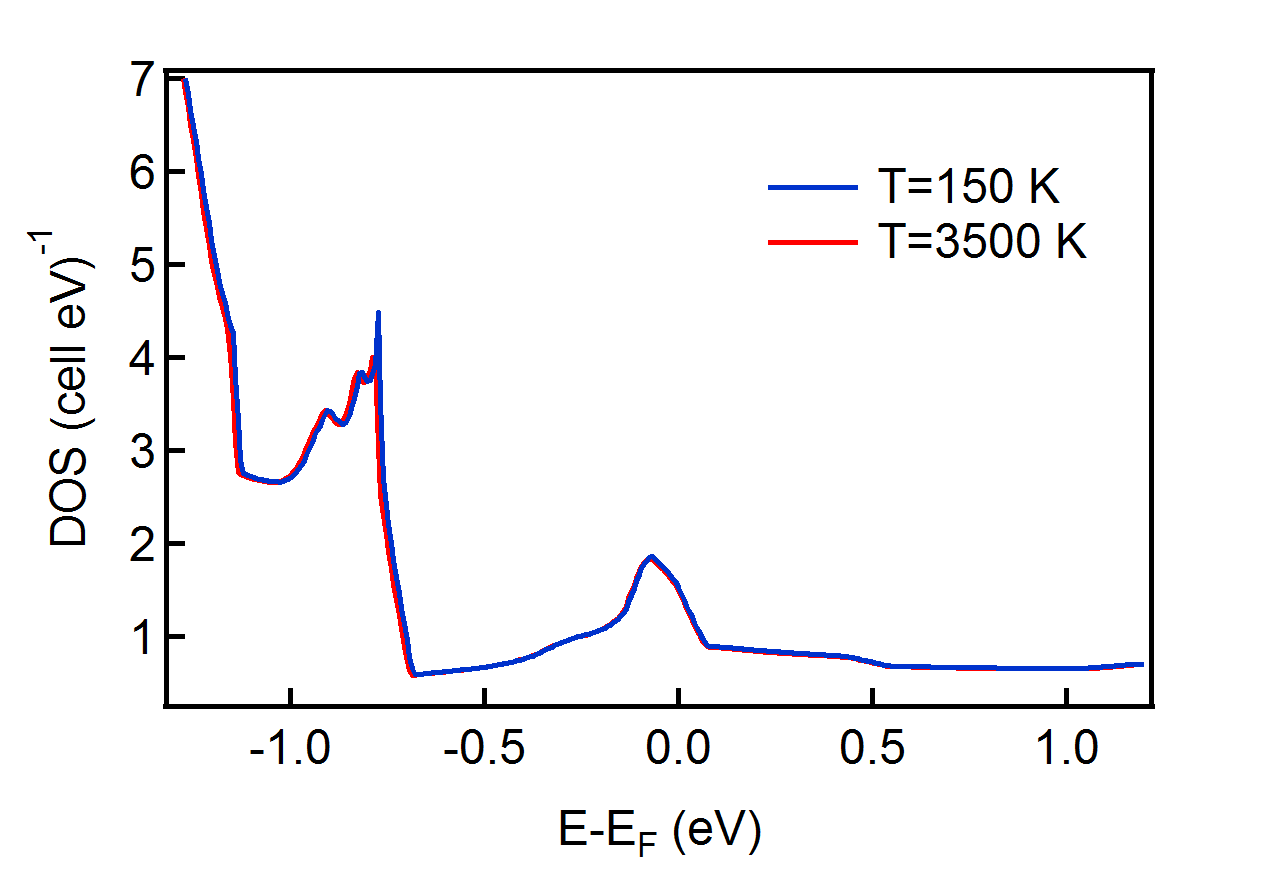}
\caption{Density of States for La$_2$CuO$_4$ self-consistently calculated at two different electronic temperatures.}
\label{DOS}
\end{figure}

The temperature-dependence of the electron-phonon coupling constant in La$_2$CuO$_4$ is calculated from the electronic structure determined using the Linear Muffin-Tin Orbital (LMTO) method in the Local Density Approximation (LDA) ~\cite{tj7}. The band structure agrees well with other band structures calculated with other methods \cite{pic89}. A single band crosses $E_F$, becomes very flat near the X-point in the Brillouin Zone and makes 
a van-Hove singularity peak in the DOS near the position of $E_F$ in undoped La$_2$CuO$_4$ \cite{tjapl09,tjprb11}.
LDA and other forms of density-functional calculations do not get the anti-ferro magnetic gap for zero doping,
but the bands describe well the electronic structure for doped cuprates, as has been verified from
ARPES~\cite{dama03}.  
Doping ($x$, in holes per Cu) is here included in a rigid-band manner to account for La/Sr substitutions. 
Spin fluctuations are neglected, and although they are certainly important for superconductivity and the low-$T$ properties of cuprates, they are quenched at the high temperatures of the experiments presented here~\cite{tj7}.

\begin{figure}
\includegraphics[width=1\linewidth,clip=true]{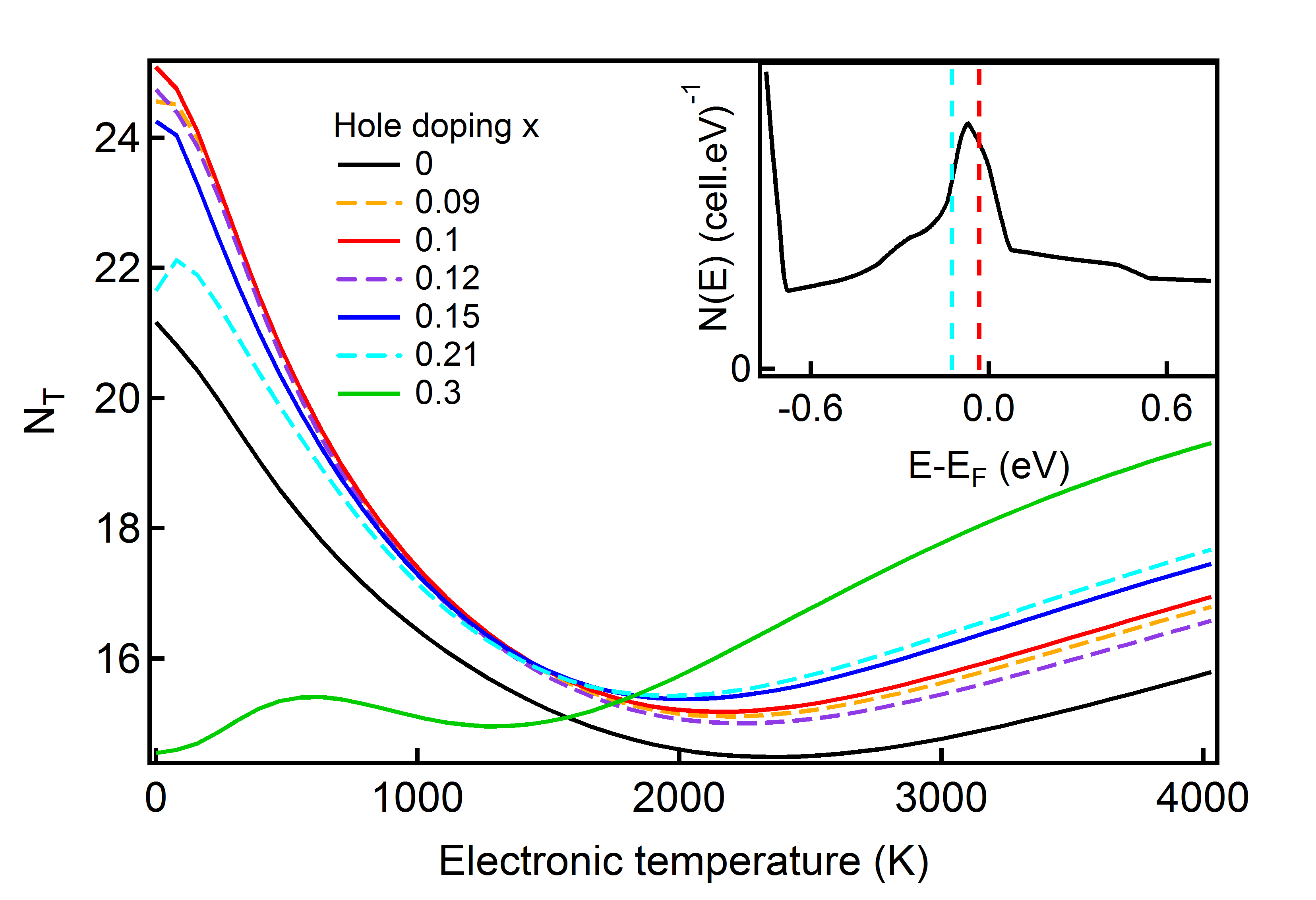}
\caption{(Color online) The $T$-dependence of the effective DOS at the chemical potential, $N_{T}$, in LSCO owing to the effect of Fermi-Dirac occupation 
for dopings $x$ indicated in the frame. Thermal disorder or disorder from lattice defects are neglected. Inset: 
The bare DOS of La$_2$CuO$_4$ near $E_F$. The vertical dashed lines indicate the rigid-band positions of $E_F$ for hole dopings 0.1 (red) and 0.21 (light blue).}
\label{fig3}
\end{figure}

A simple form \cite{zim} for the e-ph coupling constant is :
\begin{equation}
\lambda = N(E_F) B^2/M \omega^2
\label{lameq}
\end{equation}

Here, $N(E_F)$ is the DOS at $E_F$, $M$ is an atomic mass and $\omega$ a weighted average of the phonon frequency. 
The denominator is a force constant, $K = d^2E/du^2$, where $E$ is the total energy and $u$ an atomic displacement. 
The matrix element  $B = \langle \Psi^*(E_F,r) \frac{dV(r)}{du} \Psi(E_F,r)~\rangle$, can be evaluated from the band structure \cite{daco,gg,tj7}.  

The temperature dependence of $\lambda$ is mainly due to the variations of the DOS at the Fermi energy ($E_F$) produced 
by the $T$-dependence of the Fermi-Dirac occupation, $f(\varepsilon,E_F,T)=1/(\exp((\varepsilon-\mu)/(k_BT))+1))$,
where $\mu = E_F(T)$. Indeed, one could imagine that the electronic temperature could influence the DOS results if the partial DOS functions (Cu-$d$ vs O-$p$ ratios for example) vary very much within $k_B T$ around $\mu$. However, the partial DOS-ratios of this system are fairly stable within ~0.7eV from $E_F$. For energies lower than ~0.7 eV below $E_F$ (at the DOS-edge) there are some changes, but this is too far from the Fermi level to be probed by temperatures of the order 3500K. Indeed, two sets of self-consistent calculations, one at 150K and one at 3500K, produces almost identical results for the DOS (see fig.~\ref{DOS}).

Other contributions, such as disorder from thermal atomic vibrations~\cite{fesi,temp} and from lattice 
imperfections which also broaden the DOS, are neglected in the calculation of the $T$-dependent DOS $N_T$~\cite{tj84}:

\begin{equation}
N_{T}(\mu) = - \int_{-{\infty}}^{\infty} N_{eff}(\varepsilon)
\frac{\partial{f(\varepsilon,\mu,T)}}{\partial{\varepsilon}} d\varepsilon
\label{Nteq}
\end{equation}

where the effective DOS, $N_{eff}$ , can be either the bare DOS or the one calculated for a lattice with thermal disorder, or for the structure with defects. The chemical potential $\mu$ is determined from the condition of having a constant number of electrons $n$ at each $T$:

\begin{equation}
n = \int_{-{\infty}}^{\infty} N_{eff}(\varepsilon) f(\varepsilon,\mu,T) d\varepsilon
\label{neq}
\end{equation}

The $T$-variation of $N_T$, and hence the scaling of $\lambda(T)$, is shown in Fig.~\ref{fig3} as a function of the doping.
At first there is a decreasing trend of $N_{T}$ (and hence $\lambda$) for increasing $T$ since $E_F$ is close to the small van-Hove peak in the DOS (inset of Fig.~\ref{fig3}). However, the trend is reversed when $T$ is larger than $\sim$ 2000K and above, because of the beginning of high DOS feature at about 0.7 eV below $E_F$. This edge of high DOS is due to the hybridized Cu-$d$ O-$p$ bands below $E_F$ \cite{tjprb11}. 
This increase of $N_T$ will start at a lower temperature and will be stronger if structural disorders are taken into account, since the edge of the high DOS feature below $E_F$ would be smeared. Note, however, that the short pulse cannot heat the lattice during the pumping time of the experiment.
When the doping level increases, the position of $E_F$ moves to lower energy, i.e. the band edge will be 
closer to $E_F$. This explains why the $T$-dependence of $\lambda$ is stronger at large hole doping.

\section{PARTIAL ELECTRON-PHONON COUPLING CALCULATIONS}
\label{Stephen}

The partial electron-phonon couplings for each of the 21 modes of La$_{2-x}$Sr$_x$CuO$_4$ have been calculated 
using pseudopotentials, as implemented within the QUANTUM ESPRESSO code~\cite{qe}.

\begin{figure}[h!]
\includegraphics[width=1\linewidth,clip=true]{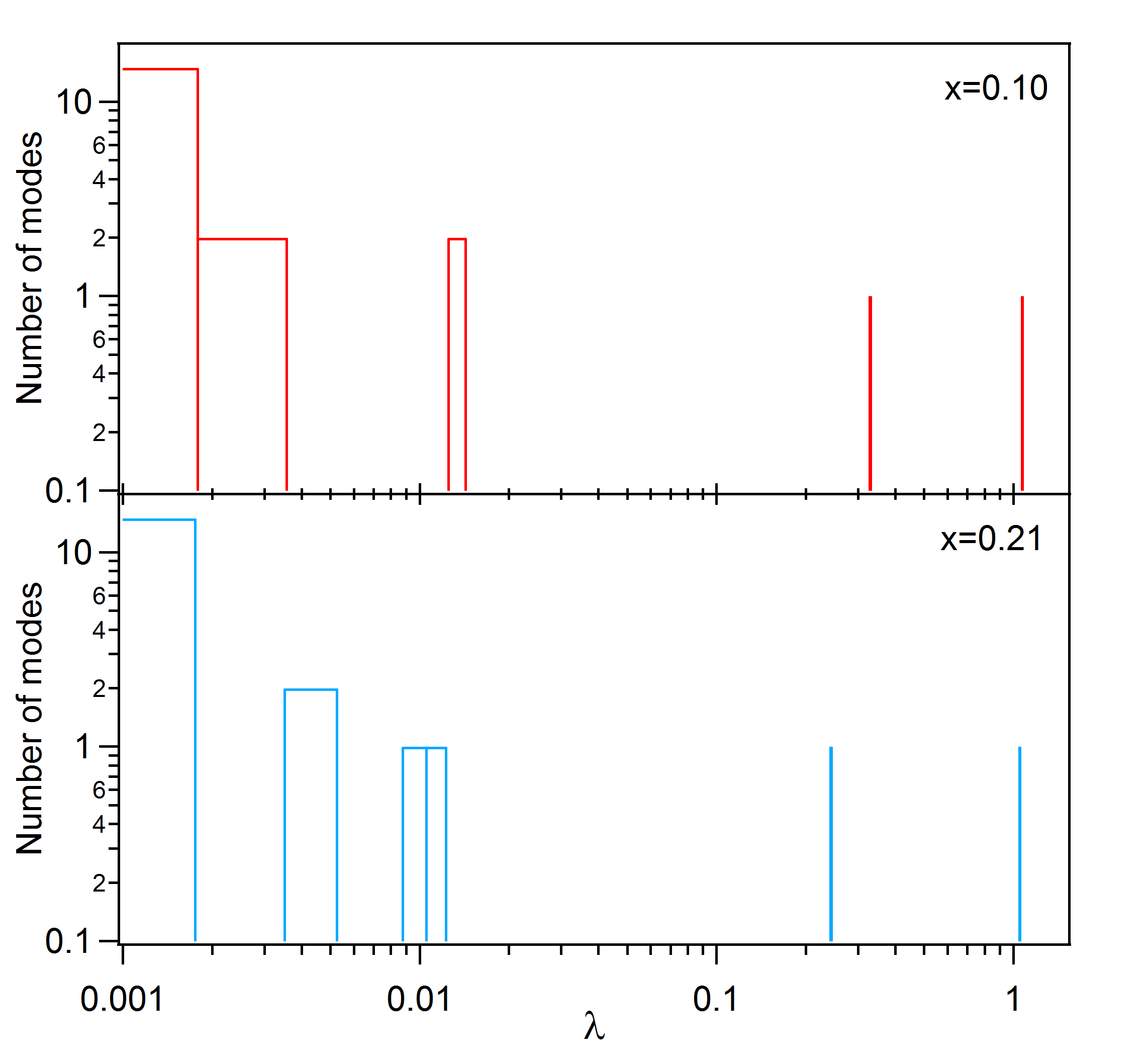}
\caption{Electron-phonon coupling constant histograms for the 21 modes at the $\Gamma$ point of 
La$_{2-x}$Sr$_x$CuO$_4$, $x=0.1$ and $x=0.21$.}
\label{lambda_histo}
\end{figure}

The results of these calculations are presented in Table~\ref{table_lambda}, and the histogram of partial electron-phonon coupling constants in Fig.~\ref{lambda_histo}. Note that the three acoustic modes are not represented in Table~\ref{table_lambda}; their $\lambda$ constants are null.

\begin{widetext}

\begin{table}[!h]
\begin {center}
\begin{tabular}{|c|c|c|c|c|c|}
\hline
mode $\sharp$ & symmetry  & E (meV) (x=0.1) & $\lambda$ (x=0.1) & E (meV) (x=0.21)  & $\lambda$ (x=0.21)\\
\hline
4 & $E_{u}$ & 4.60 & 0 & 8.98 & 0 \\
5 & $E_{u}$ & 4.60 & 0 & 8.98 & 0 \\
\hline
6 & $E_{g}$ & 7.07 & 0.0131 & 8.95 & 0.0106 \\
7 & $E_{g}$ & 7.07 & 0.0132 & 8.95 & 0.0102 \\
\hline
8 & $A_{2u}$ & 16.27 & 0 & 17.62 & 0 \\
\hline
9 & $E_{u}$ & 20.41 & 0 & 21.53 & 0 \\
10 & $E_{u}$ & 20.41 & 0 & 21.53 & 0 \\
\hline
11 & $B_{2u}$ & 21.36 & 0 & 23.91 & 0 \\
\hline
12 & $A_{2u}$ & 23.08 & 0 & 26.00 & 0 \\
\hline
13 & $A_{1g}$ & 26.86 & 0.3263 & 27.26 & 0.2398 \\
\hline
14 & $E_{g}$ & 26.93 & 0.0021 & 26.83 & 0.0042 \\
15 & $E_{g}$ & 26.93 & 0.0021 & 26.83 & 0.0040 \\
\hline
16 & $E_{u}$ & 40.89 & 0 & 42.21 & 0 \\
17 & $E_{u}$ & 40.89 & 0 & 42.21 & 0 \\
\hline
18 & $A_{1g}$ & 49.94 & 1.0587 & 50.61 & 1.0391 \\
\hline
19 & $A_{2u}$ & 57.37 & 0 & 57.76 & 0 \\
\hline
20 & $E_{u}$ & 92.36 & 0 & 94.84 & 0 \\
21 & $E_{u}$ & 92.36 & 0 & 94.84 & 0 \\
\hline
		\end{tabular}
	\caption{Symmetry, energy and partial electron-phonon coupling constant for each of the 21 modes of La$_{2-x}$Sr$_x$CuO$_4$, at the $\Gamma$ point.}
\label{table_lambda}
\end {center}
\end{table}

\end{widetext}

\begin{figure}[h!]
\includegraphics[width=0.65\linewidth,clip=true]{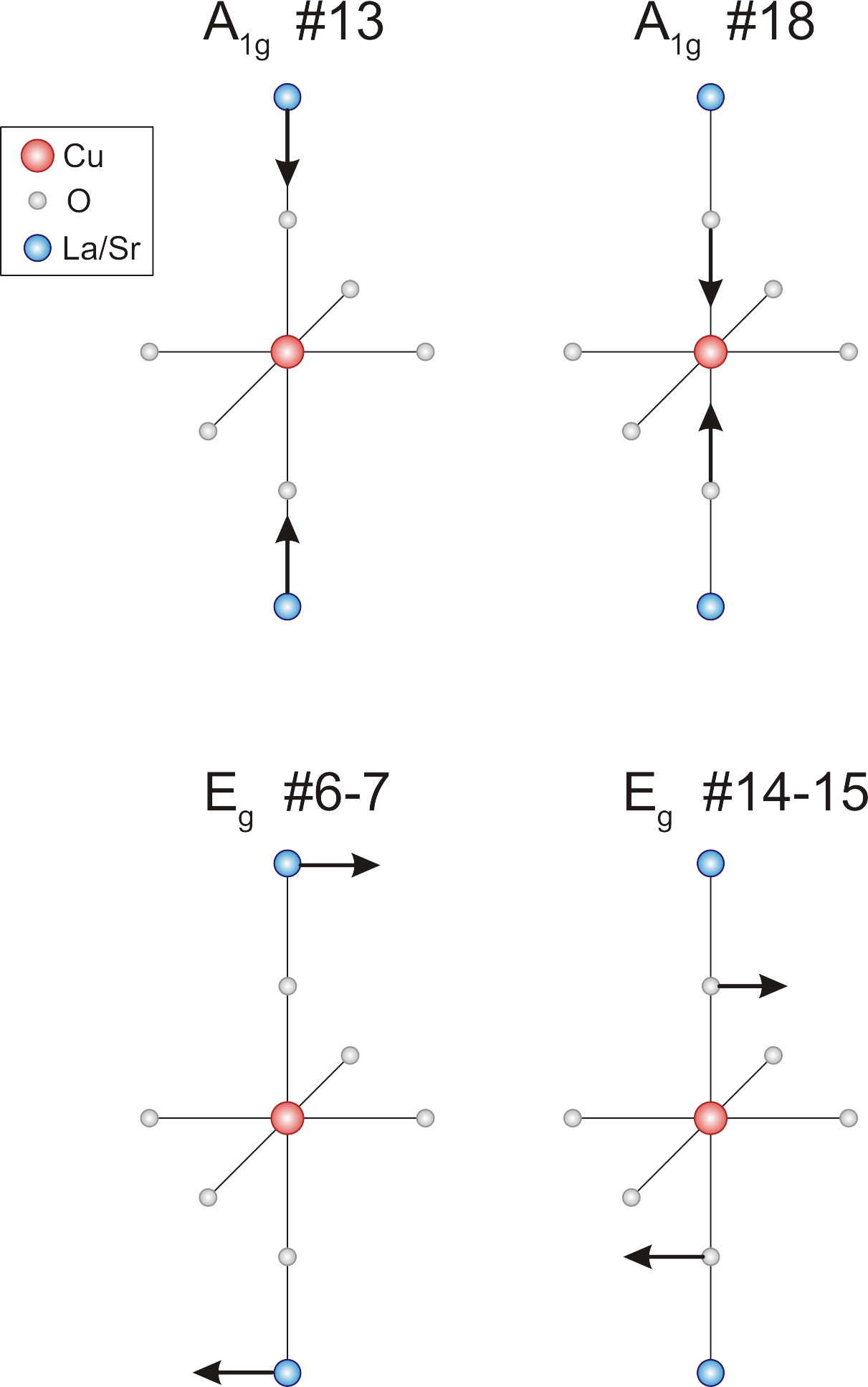}
\caption{Atomic motions corresponding to the most coupled modes in La$_{2-x}$Sr$_x$CuO$_4$, from Ref.~\cite{Mostoller1990}. Displacements smaller than $25\%$ of the maximum are not shown.}
\label{phonons}
\end{figure}

The total $\lambda$ constant, defined as the sum of all partial couplings, is $\lambda$=1.416 for $x=0.1$ and 1.308 for $x=0.21$. As expected from the density-of-states calculations at zero temperature, it is larger in the underdoped sample than in the overdoped one (see Fig. 3 of the main article). Interestingly, the two fully symmetric $A_{1g}$ modes are contributing for more than $97\%$ of the total $\lambda$; they are schematically shown in Fig.~\ref{phonons}. Since they both involve only atomic displacements along the $c$-axis, they do not influence the diffraction intensity of in-plane Bragg peaks which were detected in our time-resolved diffraction measurements. This explains why the $\lambda$ constants determined through 3TM simulations of our data are far smaller than usually predicted and measured by $k$-integrated techniques in cuprates.

Two doubly degenerate $E_g$ modes also present a finite partial $\lambda$ constant. The atomic motions involved in these modes are translation of La/Sr and apical O atoms along the $a$ or $b$-axis (see Fig.~\ref{phonons}), therefore affecting the (400) peak intensity upon excitation. As a consequence, we used the average value of their energy in order to extract $\lambda$ from $\lambda \left\langle \omega^2 \right\rangle$, the latter being obtained from the transient lattice temperature along (400), so $\left\langle \omega \right\rangle=17.32~meV$ for $x=0.1$ and $17.89~meV$ for $x=0.21$. The sum of electron-phonon coupling constants for these four modes gives $\lambda=0.031$ ($x=0.1$) and $0.029$ ($x=0.21$), in good agreement with the 3TM results at the lowest fluences, where the transient electronic temperatures are the smallest.

The `breathing'-mode, with inward-outward movements of the O-cage surrounding a Cu, is not included within a single unit cell.
Approximate results for electron-phonon coupling matrix elements in supercells extended along $x$ using the LMTO method, and 
using experimental information for the phonon energies, give $\lambda$=1.1 for planar O (displacement along $x$) and 0.13 for apical O
(along $z$) when the phonon energies are 48 and 58 meV, respectively. For La along $z$ the values are 0.02 and 17 meV, 
and the averaged $\lambda$ for the strongest modes is 0.36 \cite{tj09}. These estimates are of the same order as shown 
in Table I, but they suggest also that in-plane movements of O's can have larger $\lambda$.


\section{DISCUSSION AND CONCLUSION}

\begin{figure}[h!]
\includegraphics[width=1\linewidth,clip=true]{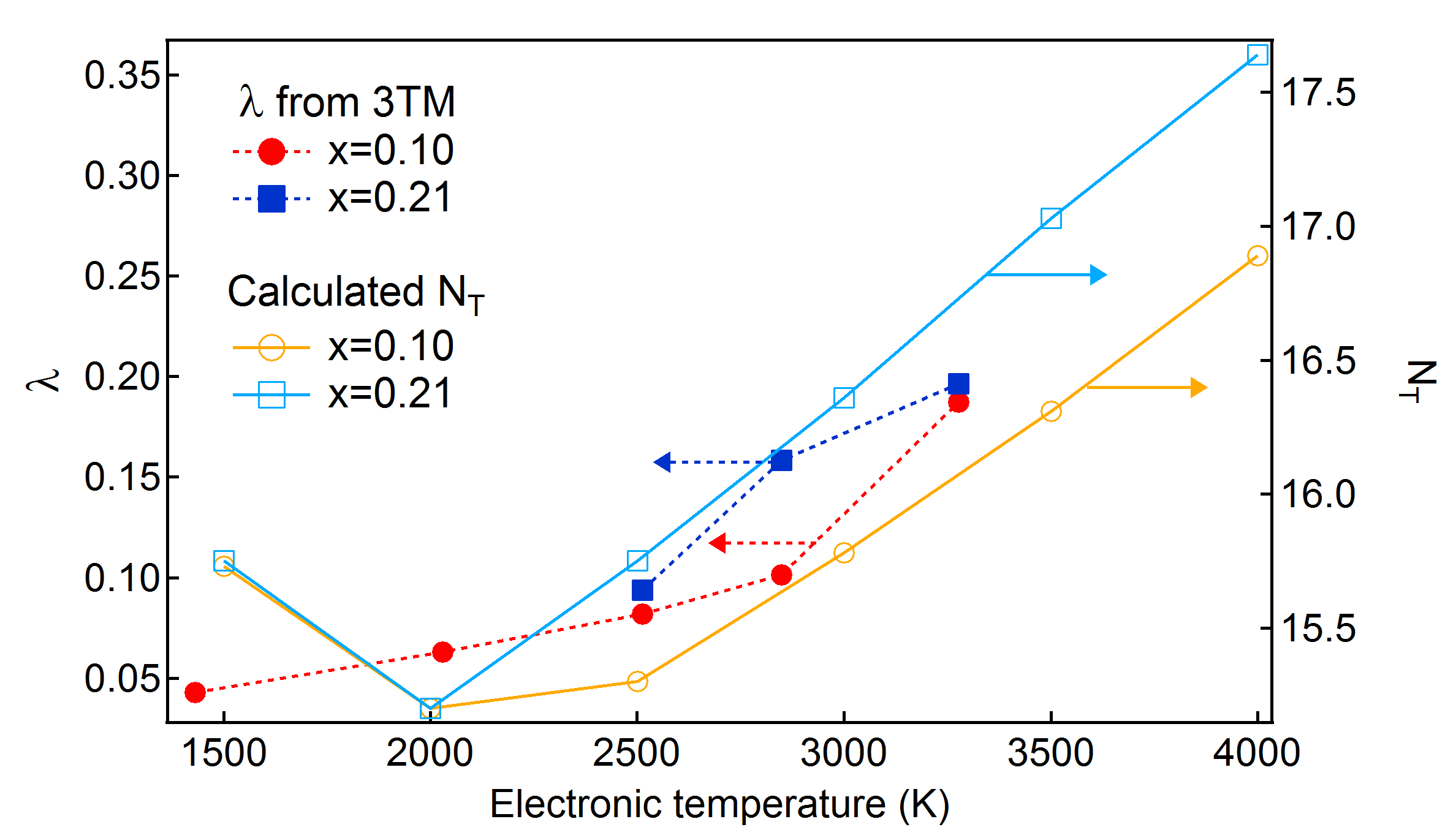}
\caption{Electron-phonon coupling constant obtained from 3TM simulations of time-resolved X-ray diffraction of the (400) peak in LSCO (solid symbols, left), and density-of-states at the Fermi level obtained by LDA calculations (empty symbols, right).}
\label{fig4}
\end{figure}

In Fig.~\ref{fig4}, the experimentally obtained e-ph coupling constants as a function of the different electronic temperatures, photoinduced in our experiments, are reported together with the values derived from the electronic structure calculations. As is clear, there is a similar trend in the temperature dependence between the experimental and calculated behavior of the e-ph coupling constant, even though we find experimentally a much stronger $T$-dependence than theoretically. This may be an effect of neglecting thermal disorder and spin fluctuations (the latter may re-appear at large $T$ in case they present a significant coupling with lattice distortions) in the calculations. Without optimized and well-tested methods for including contributions to $\lambda$ from spin-fluctuations (which can have its own $T$-dependence) in our model, we point out that one could expect a stronger $T$-dependence upon adding these excitations. Moreover, we cannot exclude the possibility that the measured peak probes a particular phonon sensitive to a part of the $\lambda$ function (the calculated DOS being a $k$-average).

The behavior of the LSCO DOS as a function of electronic temperature induces a 
temperature-dependent $\lambda$ constant. Such a non-monotonic dependence had been predicted in metals~\cite{Lin2008}, even though this would occur at much higher temperature than for LSCO. Some experimental suggestion for a $T$-dependent $\lambda$ constant was proposed in metals~\cite{Fann1992}, as well as in cuprates~\cite{Carbone2008, Carbonereview} without a clear determination of the DOS effect. In this respect, cuprates are shown to have an anomalous behavior, originated by their peculiar electronic structure. These results suggest that band effects play an important role in the electron-lattice interaction in solids, in particular for cuprate superconductors. Unveiling the evolution of these interactions throughout a larger part of the phase diagram may provide a useful feedback for the theoretical understanding of the unconventional superconductivity mechanism.

\begin{acknowledgments}
The authors thank Marisa Medarde for help during the specific heat measurements.
This work was supported by the Swiss NSF via the contracts $PP00P2-128269$ and $20020-127231/1$.
Part of this work was carried out using the computational facilities of the Advanced Computing Research Centre, University of Bristol.
\end{acknowledgments}


\begin{thebibliography}{29}
\bibitem{temp} High-T vibrations do not contribute from femtosecond excitations, since
the timescale of lattice vibrations is of the order $10^{-13}$ s.
\end{thebibliography}
\end{document}